\documentclass[11pt,a4paper]{article}
\usepackage[T1]{fontenc}
\usepackage[utf8]{inputenc}
\usepackage{lmodern}
\usepackage[english]{babel}
\usepackage{authblk}
\usepackage[a4paper,left=2.6cm,right=2.6cm,top=2.2cm,bottom=2.4cm]{geometry}
\usepackage{amsmath,amssymb,amsthm}
\usepackage{microtype}
\usepackage{xcolor}
\usepackage[colorlinks=true,linkcolor=blue!55!black,citecolor=blue!55!black,urlcolor=blue!55!black,
  pdftitle={A short proof of Seymour's Second Neighborhood Conjecture}]{hyperref}
\usepackage[nameinlink, capitalise, noabbrev]{cleveref}

\definecolor{cmt}{RGB}{176,48,32}
\newcounter{cmt}

\theoremstyle{plain}
\newtheorem{theorem}{Theorem}
\newtheorem{lemma}[theorem]{Lemma}
\newtheorem{corollary}[theorem]{Corollary}
\theoremstyle{definition}

\begin{document}

\title{A short proof of Seymour's Second Neighborhood Conjecture}
\date{}

\author[1]{Romain Bourneuf}
\author[2]{Pierre Charbit}
\author[2]{Samuel Coulomb}
\author[3]{Stéphan Thomassé}

\affil[1]{Univ. Bordeaux, CNRS, Bordeaux INP, LaBRI, UMR 5800, F-33400 Talence, France.}
\affil[2]{Université Paris Cité, CNRS, IRIF, F-75013, Paris, France.}
\affil[3]{ENS de Lyon, Université Claude Bernard Lyon 1, CNRS, Inria, LIP UMR 5668, Lyon, France.}

\maketitle

\begin{abstract}
    We give a short alternative proof of Seymour's Second Neighborhood Conjecture, recently proved by OpenAI.
    Our proof combines their reduction with a different argument showing that every nontrivial oriented graph $G$ contains a nonempty set $X$ such that $X^+\ne V(G)$ and $|(X^+)^+\setminus X^+|\geq|X^+\setminus X|$, where $X^+=N_G^+[X]$.
\end{abstract}

\section{Introduction}

In an oriented graph\footnote{An oriented graph is a directed graph with no loop and no pair of opposite arcs.} $G$, the first and second out-neighborhoods $N_G^+(v)$ and $N_G^{++}(v)$ of a vertex $v$ consist of the vertices at directed distance exactly one and two from $v$, respectively.
Seymour's Second Neighborhood Conjecture (SNC) asserts that every oriented graph contains a vertex whose second out-neighborhood is at least as large as its first.
Thus SNC is a local expansion statement: an oriented graph cannot contract at every vertex.

This conjecture was recorded by Dean and Latka~\cite{DL95}, who attributed it to Seymour and studied its restriction to tournaments\footnote{A tournament is an orientation of a complete graph.}. 
This special case, known as Dean's conjecture, was proved by Fisher~\cite{F96} using a losing probability distribution, whose existence follows from Farkas's lemma, together with an averaging argument. 
Havet and Thomassé~\cite{HT00} later gave a short combinatorial proof based on median orders.

Subsequent work approached the general conjecture in several directions.
For dense orientations, Fidler and Yuster~\cite{FY07}, together with a later correction and generalization by Ghazal~\cite{G12}, established SNC when the missing edges form a matching, a star, or a clique. 
More recently, Brukhman~\cite{B26} proved SNC whenever $n\le 2\delta^++2$, where $n$ is the number of vertices and $\delta^+$ is the minimum outdegree. 
A different line of work treated small minimum outdegree: Kaneko and Locke~\cite{KL01} settled the case $\delta^+\le 6$, and Sadhukhan and Sen~\cite{SS26} extended this to $\delta^+\le 7$.

Another natural goal was to prove an approximate form of the conjecture.
Chen, Shen and Yuster~\cite{CSY03} showed that every oriented graph $G$ has a vertex $v$ satisfying $|N_G^{++}(v)|\ge c|N_G^+(v)|$, with $c=0.657298\ldots$.
This remained the best universal constant for more than two decades, until Huang and Peng~\cite{HP24} improved it to $0.715538$.
For more background and early results on the conjecture, see Sullivan's survey~\cite[Section~3]{S06}.

Seacrest~\cite{Seacrest19} showed that SNC is equivalent to a corresponding statement for sets of vertices.
Building on this reduction, OpenAI~\cite{OAI26} strengthened the resulting set inequality and used it to prove SNC in a preprint released in October 2026.

In this paper, we give an alternative proof of SNC using a weaker form of OpenAI's inequality.

\paragraph{Proof overview.}
For a directed graph $G$ and a set $X \subseteq V(G)$, let $X^+ = N_G^+[X]$ be its closed out-neighborhood, and let $\partial^+X = N_G^+[X] \setminus X$ be its \emph{out-boundary}.
We call $b(X)=|\partial^+X| = |X^+|-|X|$ the \emph{expansion} of $X$.
Throughout, $X^+$, $\partial^+X$, and $b(X)$ are computed in the fixed ambient graph $G$.
OpenAI's strengthened inequality~\cite[Lemma~2.2]{OAI26} implies that, if SNC were false, then there would be a nontrivial oriented graph $G$ in which every nonempty proper set $X \subseteq V(G)$ satisfies 
\[
    b(X^+)<b(X).
    \tag{$\ast$}
\]

We prove that no such graph exists.

A natural approach to contradict $(\ast)$ is to start from a set $X$ and repeatedly replace it by $X^+$. 
As long as $X$ remains proper, the quantity $b(X)$ strictly decreases.
This alone gives no contradiction, since $X$ may simply grow until it reaches all of $V(G)$. 
Our idea is to run this process simultaneously from every vertex, maintaining a set $X_v$ containing $v$ for each $v\in V(G)$, and to use these sets to constrain one another so that they remain proper.

We view the family $(X_v)_{v\in V(G)}$ as a directed graph on $V(G)$: for distinct vertices $u,v$, we put an arc $u\to v$ whenever $v\in X_u$. 
We constrain the sets by requiring this directed graph to remain an oriented supergraph of $G$.
We call such a graph an \emph{extension} of $G$. 
For a directed graph $H$, we write $H_v$ for the closed out-neighborhood of $v$ in $H$. 
It is easy to see that $(\ast)$ implies that every vertex $v$ has an in-neighbor $u$ in $G$. 
Since $u\to v$ belongs to every extension $H$, orientation forces $u\notin H_v$, and hence every $H_v$ remains proper.
We start with the extension $H=G$.

Replacing every set $H_v$ by $H_v^+$ need not produce another extension, since two expanded sets may create opposite arcs. 
To avoid this, we contract the sets before expanding them. 
Given an extension $H$ of $G$, define its \emph{residual} $R$ by
\[
    R_v=\bigcap_{u\in N_G^-[v]}H_u, \qquad \text{for every $v \in V(G)$}.
\]
The definition of $R$ ensures that taking closed out-neighborhoods first in $G$ and then in $R$ reaches only vertices in the closed out-neighborhood of the starting vertex in $H$.
We write this as $GR\subseteq H$.
Observe that the residual is the largest subdigraph of $H$ satisfying $GR\subseteq H$.
The reason for this condition is simple. 
If the expanded sets $R_v^+$ created a pair of opposite arcs, their witnesses would form a closed alternating configuration of type $RGRG$. 
Read from the next vertex, this is a closed $GRGR$ configuration. 
Since $GR\subseteq H$, its two $GR$ segments form a closed $HH$ configuration. 
As $H$ is oriented, both segments must be loops. 
But every $GR$ loop is trivial: since $G,R\subseteq H$, a nontrivial one would give a digon in $H$.
Thus the original configuration is trivial, contradicting that it witnesses a digon. 
Consequently, the family $(R_v^+)_{v\in V(G)}$ defines another extension of $G$.

This suggests repeatedly replacing an extension $H$ by the corresponding extension $R^+$. 
Contraction can increase the expansion of individual sets,
so we need to control its total effect.
The quantity we hope to decrease is the \emph{total expansion}, defined as
\[
    \Phi(H)=\sum_{v\in V(G)} b(H_v).
\]
Rather than follow this process step by step, we pass directly to a terminal state: choose an extension $H$ minimizing $\Phi(H)$. 
Our main result is that, at such a minimizer, contraction cannot increase the total expansion: \[\Phi(R)\le \Phi(H).\]
Every $R_v$ is nonempty and proper, so $(\ast)$ gives
$b(R_v^+)<b(R_v)$ for every $v$. Hence
\[
    \Phi(R^+)<\Phi(R)\le \Phi(H).
\]
But $R^+$ is again an extension, contradicting the choice of $H$ as an
extension minimizing $\Phi$. This proves the theorem.

\section{Proofs}

For the contradiction, we want to show that an extension $H$ minimizing $\Phi$ satisfies $\Phi(R)\leq\Phi(H)$.
To obtain a stronger induction hypothesis, we prove this inequality under the following weaker minimality assumption.

An extension $H$ is \emph{saturated} if $\Phi(H')\ge \Phi(H)$ for every extension $H'\supseteq H$. Equivalently, for every vertex $u$, adding any set of arcs out of $u$
while preserving orientation cannot decrease $b(H_u)$.

\begin{theorem}\label{thm:residual-general}
    Let $H$ be a saturated extension of an oriented graph $G$, and let $R$ be its residual.
    Then $\Phi(R) \leq \Phi(H)$.
\end{theorem}

We prove \cref{thm:residual-general} by induction on the number of non-adjacent pairs in $H$. 
We first treat tournament extensions, which form the base case of the induction.

\begin{lemma}\label{lem:residual-tournament}
    Let $H$ be an extension of an oriented graph $G$, and let $R$ be its residual.
    If $H$ is a tournament, then $\Phi(R) \leq \Phi(H)$.
\end{lemma}

\begin{proof}
    Let $\mathcal{B}_R = \{(v,y):y\in\partial^+R_v\}$ and $\mathcal B_H=\{(v,y):y\in\partial^+H_v\}$.
    Fix $(v,y)\in\mathcal B_R$. 
    Since $H$ is a tournament, exactly one of $y\to v$ and $v\to y$ belongs to $H$.
    
    If $y\to v\in H$, then $R_v\subseteq H_v$ gives $y\in R_v^+ \subseteq H_v^+$, while $y\notin H_v$ since $H$ is oriented. 
    Hence $(v,y)\in\mathcal B_H$.

    Suppose instead that $v\to y\in H$. Since $y\in H_v\setminus R_v$, some in-neighbor $u$ of $v$ in $G$ satisfies $y\notin H_u$. 
    As $H$ is a tournament, $y\to u\in H$, and therefore $v\in\partial^+H_y$. 
    Thus $(y,v)\in\mathcal B_H$.
    
    Since $R^+$ is oriented, this gives an injection from $\mathcal B_R$ to $\mathcal B_H$, which implies \[\Phi(R)=|\mathcal B_R| \le |\mathcal B_H|=\Phi(H). \qedhere\]
\end{proof}

We can now complete the proof of \cref{thm:residual-general}.

\begin{proof}[Proof of \cref{thm:residual-general}.]
    We proceed by induction on the number of non-adjacent pairs in $H$.
    If $H$ is a tournament, the result follows from \cref{lem:residual-tournament}.
    Otherwise, choose a vertex $z$ having a non-neighbor in $H$, and let $U \subseteq V(G)\setminus\{z\}$ be the set of vertices non-adjacent to $z$ in $H$.
    Among the extensions $H'\supseteq H$ satisfying $H'_v=H_v$ for every $v\notin U$ and $z\in H'_u$ for every $u\in U$, choose one minimizing $\Phi(H')$.
    Since $H$ is saturated, the minimality of $H'$ implies that $H'$ is saturated.
    Let $R'$ denote its residual.
    Since $H'$ has fewer non-adjacent pairs than $H$, the induction hypothesis gives $\Phi(R')\leq\Phi(H')$.
    It remains to prove $\Phi(R')-\Phi(R)\geq\Phi(H')-\Phi(H)$.

    For each vertex $u\in U$, we must account for the increase $b(H'_u)-b(H_u)$ from $H$ to $H'$.
    We use Hall's theorem to associate to each $u \in U$ a distinct vertex $x_u$, such that the corresponding increase $b(R'_{x_u})-b(R_{x_u})$ from $R$ to $R'$ is at least as large.
    
    For every $S \subseteq U$, since $S$ is disjoint from $H_z$, saturation of $H$ at $z$ gives
    \[
        |S^+\setminus(H_z)^+| = |(H_z\cup S)^+|-|(H_z)^+| = |S|+b(H_z\cup S)-b(H_z) \geq |S|.
    \]
    Hall's theorem therefore gives distinct vertices
    $x_u\in N_G^+[u]\setminus(H_z)^+$ for each $u\in U$.

    For all $X, Y \subseteq V(G)$, we have $(X\cap Y)^+\subseteq X^+\cap Y^+$ and $(X\cup Y)^+=X^+\cup Y^+$, so the function $b$ is submodular: $b(X\cap Y)+b(X\cup Y)\leq b(X)+b(Y)$.

    We claim that, for every $x\in V(G)$ and every set $T$ satisfying $R_x\subseteq T\subseteq R'_x$, we have $b(R_x)\leq b(T)$.
    Indeed, choose an inclusion-minimal minimizer $T$.
    For each $y\in N_G^-[x]$, we have $T\subseteq H'_y$, so
    submodularity and saturation of $H$ at $y$ give
    \[
        b(T\cap H_y)
        \leq b(T)+b(H_y)-b(T\cup H_y)
        \leq b(T).
    \]
    Since $T\cap H_y$ remains in the interval, inclusionwise minimality forces $T\subseteq H_y$. Since this is true for all $y\in N_G^-[x]$, by definition of $R_x$ we have $T\subseteq R_x$ and hence $T=R_x$. 

    For $u\in U$, the choice of $x_u$ gives
    \[
    N_G^-[x_u]\subseteq V(G)\setminus H_z =N_H^-(z)\cup U = N_{H'}^-(z).
    \]
    Hence $z\in R'_{x_u}$.
    Since $u\in N_G^-[x_u]$, we have $R_{x_u}\subseteq H_u$ and $R'_{x_u}\subseteq H'_u$.
    The claim gives $b(R'_{x_u}\cap H_u)\geq b(R_{x_u})$.
    Since $H_u\cup\{z\}\subseteq H_u\cup R'_{x_u}\subseteq H'_u$, the minimality of $H'$ gives $b(H_u\cup R'_{x_u})\geq b(H'_u)$.
    Submodularity therefore yields
    \[
        b(R'_{x_u})+b(H_u) \geq b(R'_{x_u}\cap H_u)+b(H_u\cup R'_{x_u}) \geq b(R_{x_u})+b(H'_u),
    \]
    which rearranges to the required bound.

    By the claim, every difference $b(R'_x)-b(R_x)$ is nonnegative.
    Since the vertices $x_u$ are distinct, we obtain
    \[
    \begin{aligned}
        \Phi(R')-\Phi(R)
        &=\sum_{x\in V(G)}\bigl(b(R'_x)-b(R_x)\bigr)
          \geq\sum_{u\in U}\bigl(b(R'_{x_u})-b(R_{x_u})\bigr)\\
        &\geq\sum_{u\in U}\bigl(b(H'_u)-b(H_u)\bigr)
          =\Phi(H')-\Phi(H).
    \end{aligned}
    \]
    Together with $\Phi(R')\leq\Phi(H')$, this gives $\Phi(R)\leq\Phi(H)$.
\end{proof}

We can now derive the Second Neighborhood Conjecture.

\begin{corollary}\label{cor:snc}
    Every nonempty oriented graph contains a vertex $v$ such that $|N_G^{++}(v)|\geq |N_G^+(v)|$.
\end{corollary}

\begin{proof}
    Suppose otherwise. Then, by~\cite[Lemma~2.2]{OAI26}, there exists an oriented graph $G$ with at
    least two vertices such that $b(X^+)<b(X)$ for every nonempty proper subset $X\subseteq V(G)$.
    The graph $G$ is strongly connected: otherwise, a sink strongly connected component $X$ would be nonempty and proper and satisfy $X^+=X$, contradicting $b(X^+)<b(X)$. 
    In particular, every vertex of $G$ has an in-neighbor.
    
    Choose an extension $H$ of $G$ minimizing $\Phi(H)$, and let $R$ be its residual. 
    The extension $H$ is saturated, so \cref{thm:residual-general} gives $\Phi(R)\leq\Phi(H)$.
    Moreover, $R^+$ is an extension of $G$ by definition.

    Each $R_v$ contains $v$ and omits every in-neighbor of $v$ in $G$, since $R_v\subseteq H_v$ and $H$ is oriented.
    Thus every $R_v$ is nonempty and proper, so $b(R_v^+)<b(R_v)$.
    Summing over all vertices, we obtain
    \[
        \Phi(R^+)<\Phi(R)\leq\Phi(H).
    \]
    This contradicts the choice of $H$ as an extension of $G$ minimizing $\Phi$.
\end{proof}

Using similar arguments, we can also show that every nonempty sink-free oriented graph $G$
contains at least two vertices $v$ satisfying $|N_G^{++}(v)| \geq |N_G^+(v)|$.

\paragraph{AI statement.}
Starting from OpenAI’s proof, we explored and compared several approaches through repeated exchanges with ChatGPT Astra (OpenAI) and Claude Fable (Anthropic). Claude Fable found the proof presented here after we suggested using downward induction, with tournaments as the base case. ChatGPT Sol (OpenAI) subsequently assisted with drafting and revising the manuscript. The authors take full responsibility for the mathematical arguments, the accuracy of the references, and the final text.

\bibliographystyle{alphaurl}
\bibliography{biblio.bib}

\end{document}